\documentclass[9pt, conference, compsocconf]{IEEEtran}
\ifCLASSINFOpdf
\else
\fi
\usepackage{graphicx}
\usepackage[tight,footnotesize]{subfigure}
\usepackage[cmex10]{amsmath}
\usepackage{amsfonts}
\usepackage{amssymb}
\usepackage{textcomp}
\usepackage{bbm}        
\usepackage{algorithm}
\usepackage{algorithmic}
\usepackage{booktabs}   
\usepackage{url}
\usepackage{robinmath}

\begin{document}
%
\title{Bare Demo of IEEEtran.cls for IEEECS Conferences}
\title{Pruned Continuous Haar Transform of 2D Polygonal Patterns with Application to VLSI Layouts}


\author{\IEEEauthorblockN{Robin Scheibler}
\IEEEauthorblockA{IBM Research -- Z\"urich\\
Systems Group\\
R\"uschlikon, Switzerland\\
robin.scheibler@ieee.org}
\and
\IEEEauthorblockN{Paul Hurley}
\IEEEauthorblockA{IBM Research -- Z\"urich\\
Systems Group\\
R\"uschlikon, Switzerland\\
pah@zurich.ibm.com}
\and
\IEEEauthorblockN{Amina Chebira}
\IEEEauthorblockA{Swiss Federal Institute of Technology\\
Audiovisual Communications Laboratory\\
Lausanne, Switzerland\\
amina.chebira@epfl.ch}
}


%


\maketitle

\begin{abstract}
We introduce an algorithm for the efficient computation of the continuous Haar
transform of 2D patterns that can be described by polygons.  These patterns are
ubiquitous in VLSI processes where they are used to describe design and mask
layouts. There, speed is of paramount importance due to the magnitude of the
problems to be solved and hence very fast algorithms are needed. We show that
by techniques borrowed from computational geometry we are not only able to
compute the continuous Haar transform directly, but also to do it quickly. This
is achieved by massively pruning the transform tree and thus dramatically
decreasing the computational load when the number of vertices is small, as is
the case for VLSI layouts. We call this new algorithm the pruned continuous
Haar transform. We implement this algorithm and show that for patterns found in
VLSI layouts the proposed algorithm was in the worst case as fast as its
discrete counterpart and up to 12 times faster.
\end{abstract}


%
\IEEEpeerreviewmaketitle

\section{Introduction}
\label{sec:intro}

The Haar transform (HT) is often a tool of choice in image processing due to
its edge detection property, low complexity and the simplicity of its
implementation. It is particularly suited for piecewise constant functions that
have a very sparse and accurate representation in the Haar domain. An important
class of two-dimensional (2D) piecewise constant functions is the class of
functions described by a union of disjoint polygonal subsets of $\mathbb{R}^2$.
Such a description is often used in different areas of image processing such as
contour detection, segmentation, tomography image reconstruction
\cite{milanfar_reconstructing_1995} or for VLSI layouts description
\cite{semi_oasis_2008}. A polygonal shape is usually described by an ordered
list of its vertices. This description has the advantage of being very compact
and natural to understand.  Many algorithms in computational geometry make
efficient use of this description to solve various problems like intersections
of polygons, area computations or point inclusion
\cite{preparata_computational_1985}. However, this description has no fixed
length which makes it more cumbersome for use in other applications in image
processing including machine learning, pattern matching or measuring
similarity. The Haar transform provides such a fixed-length representation.

Optical lithography is the process that allows mass production of VLSI circuits
\cite{mack_fundamental_2008}. The HT has been used so far in lithography to
compress the Fourier precompensation filters for electron beam lithography
\cite{haslam_two-dimensional_1985} and also to regularize the obtained mask in
inverse lithography \cite{ma_generalized_2007}. More recently, Kryszczuk et al.
introduced the direct printability prediction of VLSI layouts using machine
learning techniques \cite{kryszczuk_direct_2010}. They use fixed-length feature
vectors from orthogonal transforms and train a classifier to predict the
printability of VLSI layouts without having to go through the detailed, and
thus computationally expensive, simulation of the physical process of
lithography. The HT is a perfect candidate to provide features in that case due
to its close match to the polygons found in VLSI layouts. However, to obtain
these features one has first to perform the transform of the enormous amount of
data contained in modern VLSI layouts. It is thus crucial to have a very fast
algorithm to yield HT coefficients from the vertex description of the polygons.

The most straightforward way would be to first create a discrete image
by sampling the polygons, and then use the discrete Haar transform (DHT) on the
resulting image. However, the polygons describe an inherently continuous
function, which allows us to compute the continuous Haar transform (CHT)
coefficients instead. By using techniques borrowed from computational geometry
to compute the inner products with the CHT basis functions, we are able to
massively prune the transform flow-diagram in addition to avoiding sampling
completely. This leads to a dramatic decrease of the computational load when
the number of vertices is small. We call this new algorithm pruned continuous
Haar transform (PCHT). The outputs of the DHT and the PCHT are identical for 2D
polygonal patterns. The PCHT was concretely implemented in a lithography tool
and proved to have significantly lower runtime compared to the DHT for the
particular case of rectilinear polygons from VLSI design layouts.

The main contribution of this work is PCHT, a fast algorithm, and to the best
of our knowledge the first of its kind for the CHT of 2D piecewise constant
polygonal patterns. Its efficiency compared to the DHT for the case of VLSI
layouts is demonstrated with potential high impact for pattern matching
techniques envisioned in computational lithography.

In \sref{polygons}, we will first briefly introduce the signal model we
consider, namely 2D piecewise constant polygonal patterns. In \sref{haar}, we
give a reminder on the 1D and 2D Haar transform while \sref{algo} presents the
PCHT algorithm. The results of the application to VLSI design layouts are given
in \sref{appli} and we conclude in \sref{concl}.


\section{Signal Model}
\slabel{polygons}

The signal model we consider is one of 2D piecewise constant polygonal
patterns. This is the class of images described by the union of a finite number
of disjoint \emph{simple polygons}. Mathematically, a polygon is described by a set of
points called vertices. A polygon $\Pol\subset\R^2$ is typically defined by its
boundary, which is a collection of straight segments, called edges. The polygon
contains all the points inside the boundary. The description of a polygon is
the list of its $K$ vertices, ordered clockwise:
\begin{equation}
\left\{(x_0,y_0), \ldots, (x_{K-1},y_{K-1})\right\}, \qquad (x_i, y_i) \in \R^2.
\nonumber
\end{equation}
where any two successive vertices describe one edge of the boundary. In
addition, simple polygons have the property that no two edges intersect each
other. A subclass of simple polygons termed rectilinear comprises all those
with only right angles and is the building block of VLSI layouts. An example of
such polygons is shown in \ffref{polygons}.

\begin{figure}[tb]
  \centering
  \centerline{\includegraphics[width=\linewidth]{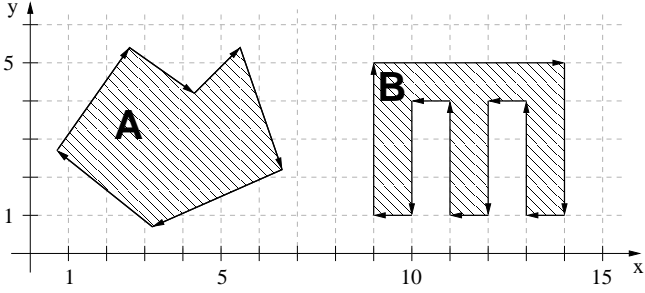}}
  \caption{Example of polygons. Polygon A is a simple polygon. Polygon B is a rectilinear polygon such as those found in VLSI layouts.}
  \flabel{polygons}
\end{figure}

A 2D piecewise constant polygonal \emph{pattern} is described by a collection of $M$
disjoint polygons each with an associated weight
$\left\{\left(\Pol_i,w_i\right)\right\}_{i=0}^{M-1}$ and with disjoint interiors,
i.e. $\Int\left(\Pol_i\right) \cap \Int\left(\Pol_j\right) =
\emptyset$ $\forall i\neq j$, where $\Int(\Pol)$ is the interior of polygon
$\Pol$, and $w_i\in\R$. Finally the continuous image model is
\begin{equation}
f(x,y) = \sum\limits_{i=0}^{M-1}w_i \indic{\Pol_i}(x,y)
\elabel{sig_model}
\end{equation}
where we use an indicator function $\indic{\Pol}(x,y) = 1$ if $(x,y)\in\Pol$
and 0 otherwise.

\section{The Haar Transform}
\slabel{haar}

The 1D Haar basis is an orthonormal basis on $[0,T)$, composed of the family of functions 
$$\left\{\varphi^{(T)}_{0,0},\psi^{(T)}_{j,k} \Big|\, j\in\N,\,k=0,\ldots,2^j-1\right\}$$
where
\begin{equation}
\varphi^{(T)}_{j,k}(t) = \frac{2^{\frac{j}{2}}}{\sqrt{T}} \varphi\left(\frac{2^j}{T}t-k\right),
\quad
\psi^{(T)}_{j,k}(t) = \frac{2^{\frac{j}{2}}}{\sqrt{T}} \psi\left(\frac{2^j}{T}t-k\right).
\nonumber
\end{equation}
The functions $\varphi$ and $\psi$ are respectively defined as
\begin{equation}
\varphi(t) =
\begin{cases}
  1 & \text{if $0 \leq x < 1$} \\
  0 & \text{otherwise}
\end{cases}
\nonumber
,\ 
\psi(t) =
\left\{
\begin{array}{rl}
  1 & \text{if $0 \leq t < \frac{1}{2}$} \\
 -1 & \text{if $\frac{1}{2} \leq t < 1$} \\
  0 & \text{otherwise}
\end{array}.
\right.
\nonumber
\end{equation}

As the 2D Haar basis is separable, we can define it in terms of the 1D basis.
Now we want to work over a surface $\fat{T} = [0,T_x) \times [0,T_y)$ we call a
tile. The scaling function is given by:
\begin{equation}
\varphi_{j,k_x,k_y}(x,y) 
  = \varphi^{\left(T_x\right)}_{j,k_x}(x) \varphi^{\left(T_y\right)}_{j,k_y}(y),
\elabel{ll}
\end{equation}
where $j=k_x=k_y=0$. The other basis functions are given by the possible
combinations of $\varphi$ and $\psi$, one of them being:
\begin{equation}
\psi^{(hl)}_{j,k_x,k_y}(x,y) = \psi^{\left( T_x \right)}_{j,k_x}(x) \varphi^{\left( T_y \right)}_{j,k_y}(y),
\elabel{hl}
\end{equation}
where $j \in \mathbb{N}$ is the scale and $k_x,\,k_y \in \{0, \ldots, 2^j-1\}$
the shifts in the $x$ and $y$ directions respectively. Similarly, we get
$\psi^{(lh)}_{j,k_x,k_y}$ and $\psi^{(hh)}_{j,k_x,k_y}$. The first and second
letter in the superscript indicate which basis function is used for $x$ and $y$
directions respectively, $h$ and $l$ indicate $\psi$ and $\varphi$
respectively.

Given the basis functions defined in \eref{ll} and \eref{hl}, we derive the
Haar transform as the inner product between the function $f$ to transform and
the Haar basis functions. Using the $L^2(\fat{T})$ and the
$l^2(\hat{\vect{T}})$ inner products, with a discretized tile $\hat{\fat{T}}$
and discretized basis functions, we respectively get the dyadic continuous and
discrete transforms \cite{vetterli_world_2009}. The CHT and DHT coefficients
are identical for 2D piecewise constant polygonal patterns.

Both the CHT and the DHT can be computed using the fast orthogonal wavelet
transform (FWT) \cite{mallat_wavelet_2008}. This algorithm is constructed using
the two-scale relationships that link the basis functions at different scales:
\begin{equation}
\varphi(t) = \sqrt{2} \sum\limits_n g_n \varphi(2t-n),\quad \psi(t) = \sqrt{2} \sum\limits_n h_n \varphi(2t-n),
\nonumber
\end{equation}
where $g_n$ and $h_n$ are the taps of two discrete-time filters
\cite{vetterli_world_2009}. The Haar filters are defined as $g_n = [ 2^{-1/2}
\; 2^{-1/2} ]$ and $h_n = [ 2^{-1/2} \;-2^{-1/2} ]$. This results in a
Cooley-Tukey butterfly structure \cite{ahmed_orthogonal_1975} where only the
inner products with the scaling function at the lowest level need be computed.
The full flow diagram for a length-8 1D transform is shown in light grey in
\ffref{sigflow}.

Using the separability of the 2D transform and the two-scale relationships, we
obtain the relations between the 2D basis functions of the different scales.
For example, $\psi^{(hl)}_{j,k_x,k_y}$ can be written as
\begin{equation}
\psi^{(hl)}_{j,k_x,k_y}(x,y) = \sum\limits_n \sum\limits_m h_n g_m \varphi_{j+1,2k_x+n,2k_y+m}(x,y). \nonumber
\end{equation}
By replacing $h_ng_m$ in the sum by $g_ng_m$, $g_nh_m$ and $h_nh_m$ we obtain
$\varphi_{j,k_x,k_y}$, $\psi^{(lh)}_{j,k_x,k_y}$ and $\psi^{(hh)}_{j,k_x,k_y}$
respectively. As in the 1D case, these relations induce a 2D butterfly
structure. Therefore transform coefficients can be computed as a linear
combination of inner products with the scaling function at different scales. 

\section{Pruned Continuous Haar Transform}
\slabel{algo}

\begin{figure}[tb]
  \centering
  \centerline{\includegraphics[width=\linewidth]{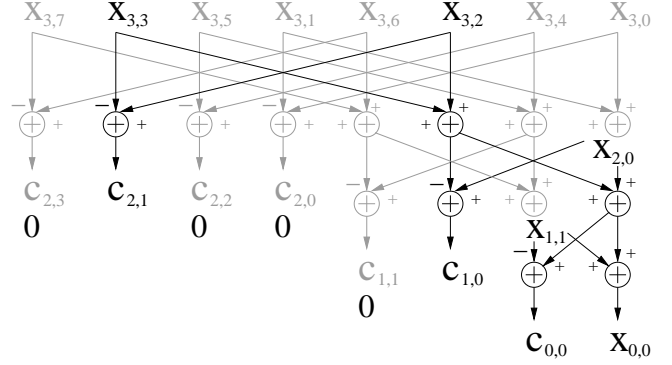}}
  \caption{A pruned signal flow of the 1D CHT. The full flow-diagram is
           shown in light grey. The signal transformed is $f(t) = u(t-3)$, 
           defined on $[0,8)$, where $u(t)$ is the Heaviside function.
           $X_{j,k} = \ip{f}{\varphi_{j,k}}$ and $C_{j,k} = \ip{f}{\psi_{j,k}}$.}
  \flabel{sigflow}
\end{figure}

\subsection{Algorithm Derivation}

Let us consider the FWT described in \sref{haar}. First, using the signal
model from \eref{sig_model} and the linearity of the inner product, we can
decompose the transform into a sum of inner products of individual polygons
with basis functions:
\begin{equation}
  \ip{f}{\varphi_{j,k_x,k_y}} = \sum\limits_{i=0}^{M-1} w_i \ip{\indic{\Pol_i}}{\varphi_{j,k_x,k_y}}. \nonumber
\end{equation}
Thus, from now on we consider only the transform of a single polygon. The
second idea is to use computational geometry techniques to compute the inner
product. The continuous inner product between the indicator of a polygon and
the scaling function is the area of the geometrical intersection of the polygon
and the support of the scaling function, multiplied by $2^j/\sqrt{T_x T_y}$.

The Haar transform acts as a discontinuity detector and all the transform
coefficients will be zero except for basis functions that intersect the
boundary of the polygon.  As a consequence, the basis functions completely
inside or outside a polygon yield a zero inner product. Moreover, all the
coefficients below such a basis function in the transform tree are also zero
(see \ffref{sigflow}). Therefore the transform can be written as a
divide-and-conquer algorithm. Divide the tile in four rectangular parts
recursively until the part considered is completely inside or outside the
polygon. Pseudocode for the PCHT is given in \algref{transform}, in which
$\fat{T}_{j,k_x,k_y} =
[k_xT_x/2^j,(k_x+1)T_y/2^j)\times[k_yT_y/2^j,(k_y+1)T_y/2^j)$ is the support of
$\varphi_{j,k_x,k_y}$. An example of the pruned transform flow-diagram for the
1D case is shown in black in \ffref{sigflow}. In order to compute all the
transform coefficients, the algorithm is called in the following way:
\begin{equation}
X_{0,0,0} = s_0\sum\limits_{i=0}^{M-1} w_i \operatorname{PCHT}\left(\Pol_i, 0, 0, 0, s_0 w_i\right), \nonumber
\end{equation} 
where $X_{j,k_x,k_y} = \ip{f}{\varphi_{j,k_x,k_y}}$ and $s_0=1/\sqrt{T_xT_y}$. 
The transform coefficients are $C^{(ab)}_{j,k_x,k_y} = \ip{f}{\psi^{(ab)}_{j,k_x,k_y}}$.

\begin{algorithm}[t]
  \caption{PCHT$(\Pol,j,k_x,k_y,s)$} \alabel{transform}
    \begin{algorithmic}[1] 
    \REQUIRE The polygon to transform $\Pol$, the scale $j$, the shifts $k_x$ and $k_y$ and the scaling factor $s$.
    \ENSURE $C^{(hl)}$,$C^{(lh)}$,$C^{(hh)}$ contain the transform coefficients.

    \STATE $i \gets \text{IntersectionArea}\left(\Pol,\fat{T}_{j,k_x,k_y}\right)$
    \IF{$i=0$ or $i=T_xT_y/2^{2j}$ or $j=J$}
      \STATE Return $i$
    \ENDIF

    \STATE $x \gets \text{PCHT}(\Pol,j+1,2k_x,2k_y,2s)$
    \STATE $y \gets \text{PCHT}(\Pol,j+1,2k_x+1,2k_y,2s)$
    \STATE $z \gets \text{PCHT}(\Pol,j+1,2k_x,2k_y+1,2s)$
    \STATE $t \gets \text{PCHT}(\Pol,j+1,2k_x+1,2k_y+1,2s)$

    \STATE $a \gets x-y$
    \STATE $b \gets x+y$
    \STATE $c \gets z-t$
    \STATE $d \gets z+t$

    \STATE $C^{(hl)}_{j,k_x,k_y} \gets C^{(hl)}_{j,k_x,k_y} + s(a+c)$
    \STATE $C^{(lh)}_{j,k_x,k_y} \gets C^{(lh)}_{j,k_x,k_y} + s(b-d)$
    \STATE $C^{(hh)}_{j,k_x,k_y} \gets C^{(hh)}_{j,k_x,k_y} + s(a-c)$

    \STATE Return $b+d$

    \end{algorithmic}
\end{algorithm}

\begin{figure}[tb]
  \centering
  \centerline{
    \begin{minipage}{0.28\linewidth}
      \centering
      \centerline{\includegraphics[width=\linewidth]{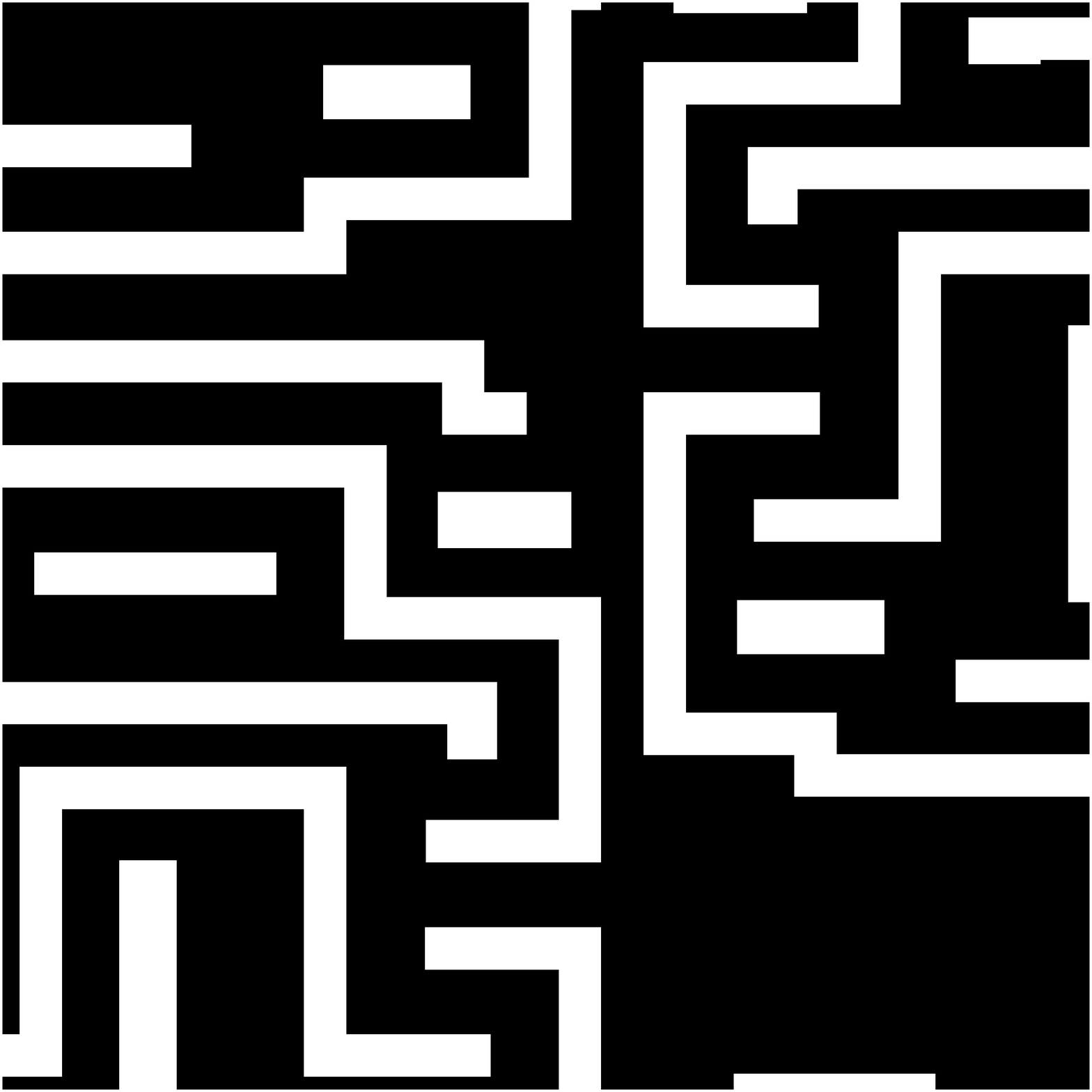}}
    \end{minipage}
    \hfill
    \begin{minipage}{0.28\linewidth}
      \centering
      \centerline{\includegraphics[width=\linewidth]{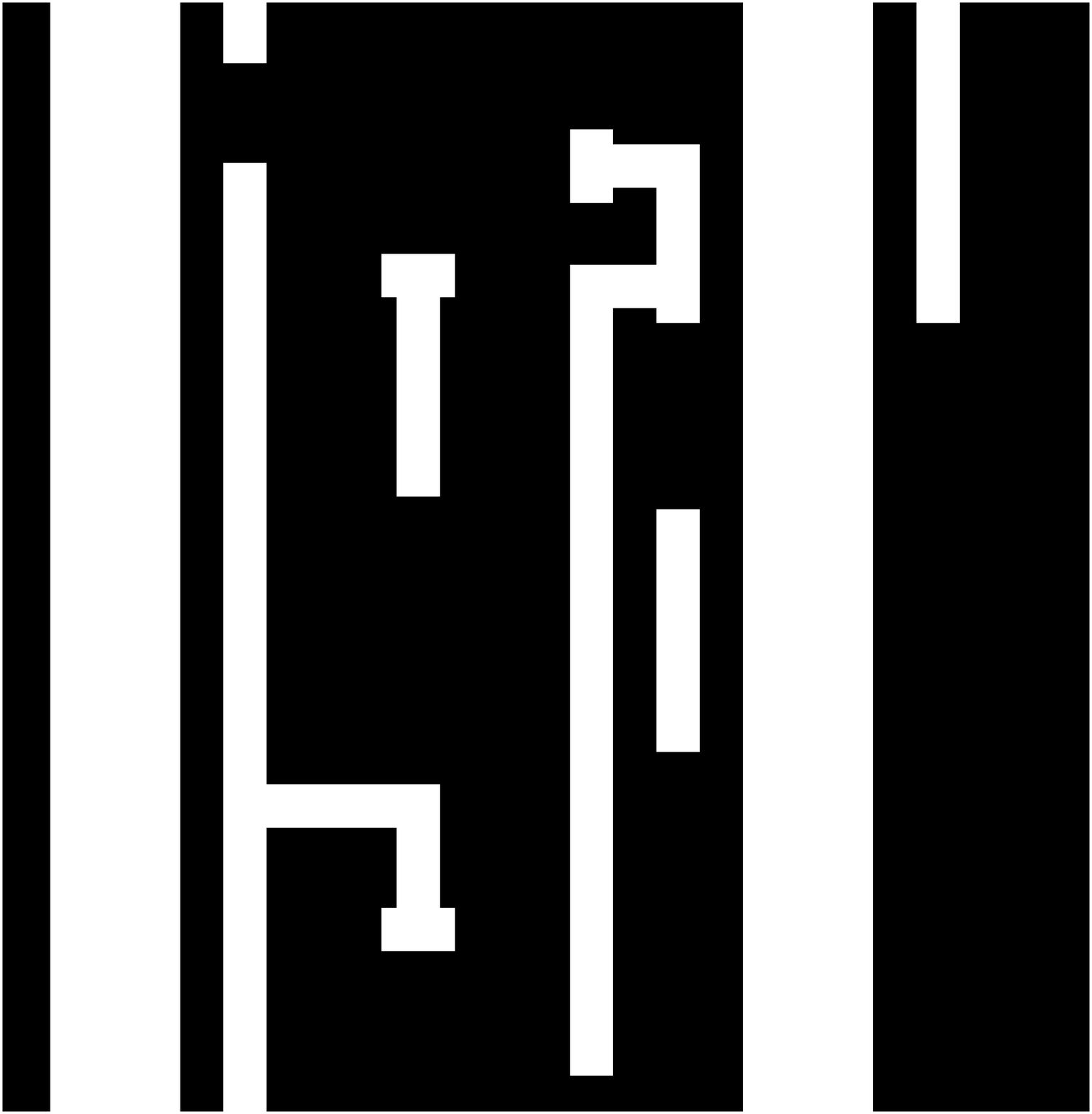}}
    \end{minipage}
    \hfill
    \begin{minipage}{0.28\linewidth}
      \centering
      \centerline{\includegraphics[width=\linewidth]{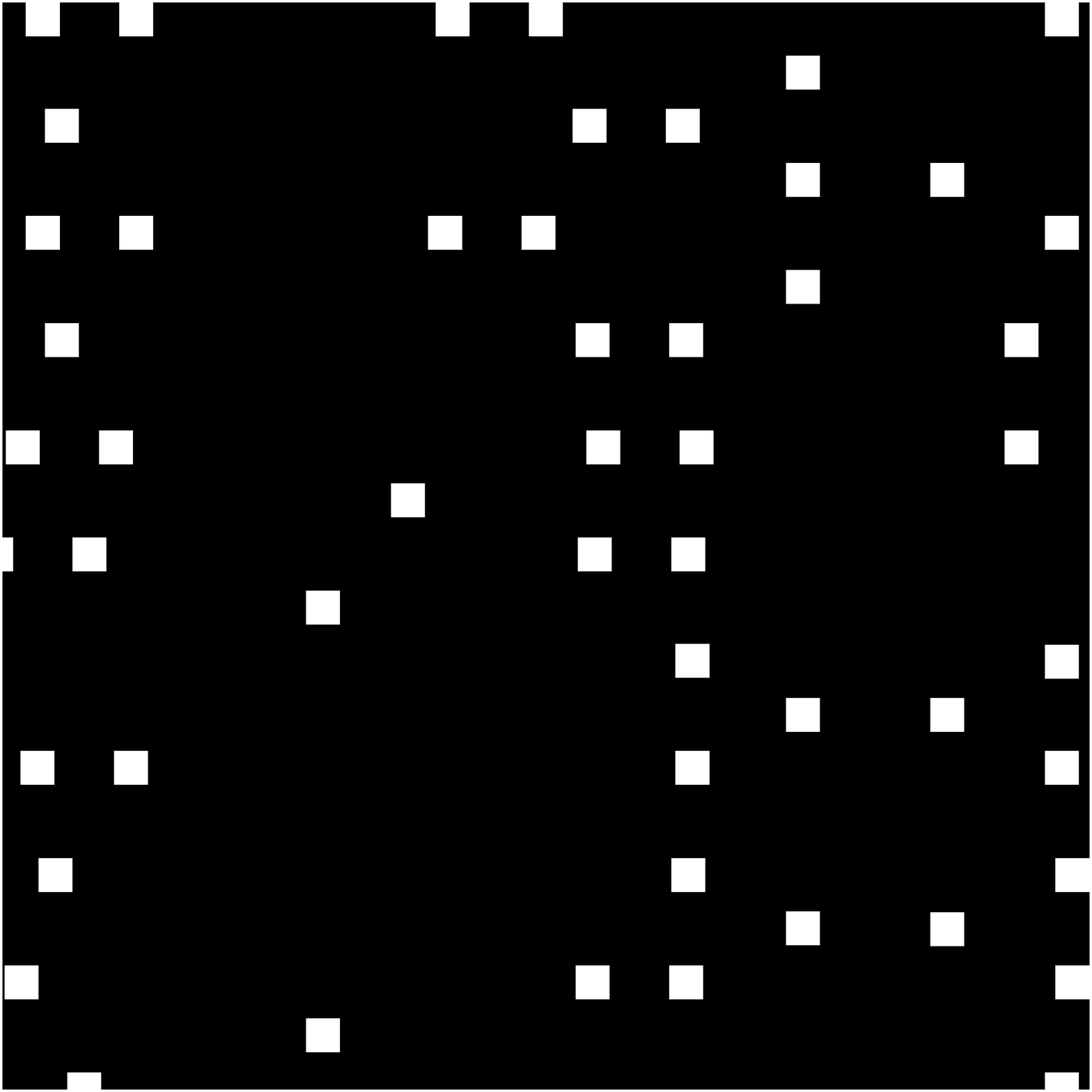}}
    \end{minipage}
  }
  \vspace{10pt}
  \caption{From left to right: Examples of 1024nm$\times$1024nm tiles from respectively M1, M2 and CA layers.}
  \flabel{layout}
\end{figure}

\section{Application to VLSI Layouts}
\slabel{appli}

We now show a practical example of the application of this algorithm to compute
the CHT coefficients of a VLSI layout. In practice a layout is described using
a vector format such as OASIS \cite{semi_oasis_2008}. Compared to general
simple polygons, those found in VLSI layouts have two additional properties.
All vertices are placed on the integer grid and all edges are parallel to
either the $x$ or $y$ axis. We call these polygons rectilinear. The routine
used to compute the intersection area in \algref{transform} is specifically
adapted for rectilinear polygons from classical computational geometry
techniques \cite{preparata_computational_1985}.

We implemented the PCHT and DHT algorithms in a computational lithography tool
that we ran on a 3GHz Intel Xeon 5450 running Linux in 64-bit mode. All the
code is C++, single-threaded and was compiled using GCC 4.1.2 with option
``\mbox{-O3}''. The DHT was custom implemented taking into account the knowledge that
the input image is binary. For both transforms the output coefficients were
discarded instead of being stored in order to minimized the impact of memory
transfers on the runtime measurements. We ran a benchmark of the PCHT and the
DHT on different layers of a 22nm VLSI layout. Layers M1 and M2 are metal
layers that contain both rectangles and other polygons, while the contact array
(CA) layer contains only rectangles. Examples of tiles from the different
layers are shown in \ffref{layout}. \ffref{K_vs_RT} shows the runtime as a
function of the number of vertices $K$ in 1024nm$\times$1024nm tiles from the
M1 layer, the worst configuration for the PCHT in our experiment. The empirical
distribution of the number of vertices in the tiles is shown in light grey.
Although the runtime of the continuous transform grows with $K,$ it outperforms
its discrete counterpart for about half the tiles. The peaks around $K=190$ are
caused by the very low number of tiles in that range, as shown by the empirical
distribution, and all these tiles having a worse than average runtime. The DHT
also shows a slight dependence on $K$ due to the time needed to create the
discrete image. The average speed-up of the runtime as a function of the tile
size is shown in \ffref{N_vs_SU}. The speed-up is defined as the ratio of the
runtimes of the DHT and pruned CHT. Because it has the highest vertex density,
the M1 layer shows the least improvement, between 1 and 3 times speed-up. For
large tiles, layers M2 and CA show speed-up over 6 and 12 times, respectively.
In a practical scenario where the output coefficients need to be stored for
further use we expect the PCHT to outperform even more the DHT as its pruned
structure avoids completely the computation of zero coefficients and thus their
storage and associated memory transfers. On the other hand, the DHT has no
knowledge of which coefficients will be zero and thus either tries to store
every output coefficients or an if statement can be used if we know in advance
that most coefficients will be zero, as is the case here. In addition, further
optimizations such as parallelization and optimization for the cache size are
possible.

\section{Conclusions}
\slabel{concl}

\begin{figure}[tb]
  \centering
  \centerline{
    \begin{minipage}{\linewidth}
      Median Runtime [ms] \\
      \centerline{\includegraphics[width=\linewidth]{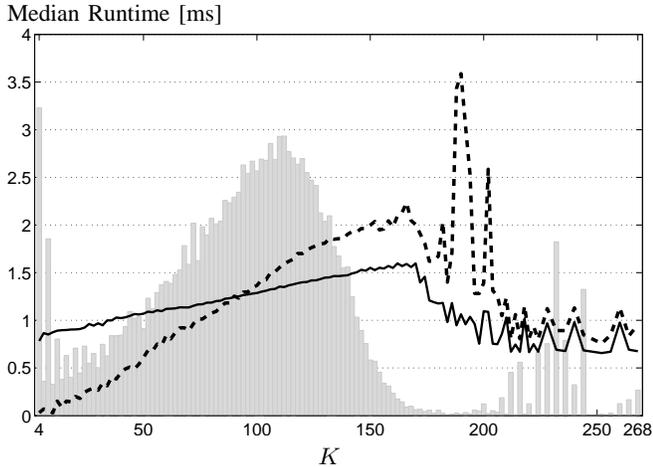}} \\
      \centerline{$K$} \\[-5pt]
    \end{minipage}
  }
  \caption{Median runtime of the PCHT (dashed line) and DHT (plain line) of
          1024nm$\times$1024nm tiles from the M1 layer containing $K$ vertices.
          Tiles with $K>200$ have a lower runtime because they contain exclusively
          rectangles which are less complex. The empirical distribution of the number of
          vertices is shown in grey.}
  \flabel{K_vs_RT}
\end{figure}

We introduced the PCHT, a new algorithm for the computation of the CHT of 2D
polygonal patterns. We showed significant speed-up compared to the DHT in an
implementation targeting rectilinear polygons found in VLSI layouts. We expect
the PCHT to impact machine learning techniques being developed for application
in computational lithography, such as printability prediction
\cite{kryszczuk_direct_2010}, as well as in the VLSI design process in general.

The natural next step for our work would be to analyze the computational
complexity of PCHT in order to validate theoretically its superiority over the
DHT. The performance of both algorithms implementations should also be
reassessed when the coefficients are stored in memory in order to account for
the impact of memory transfers. Another important implementation step is the
parallelization of the code as current and future increases in computation power
come primarily through multi-core chips. The recursive nature of the PCHT
algorithm makes it a perfect candidate for parallelization.

We also intend to apply our vertex based approach to other transforms such as
the continuous Fourier series. A fast algorithm to compute the Fourier series
would also be very valuable in lithography where the fast Fourier transform is
routinely used in optical lithography process simulation despite the
introduction of aliasing due to the infinite bandwith of the polygons.
Moreover, we expect an optimized vertex based fast continuous Fourier series
algorithm to be inherently faster than the fast Fourier transform.

\begin{figure}[tb]
  \centering
  \centerline{
    \begin{minipage}{\linewidth}
      Speed-up \\[3pt]
      \centerline{\includegraphics[width=\linewidth]{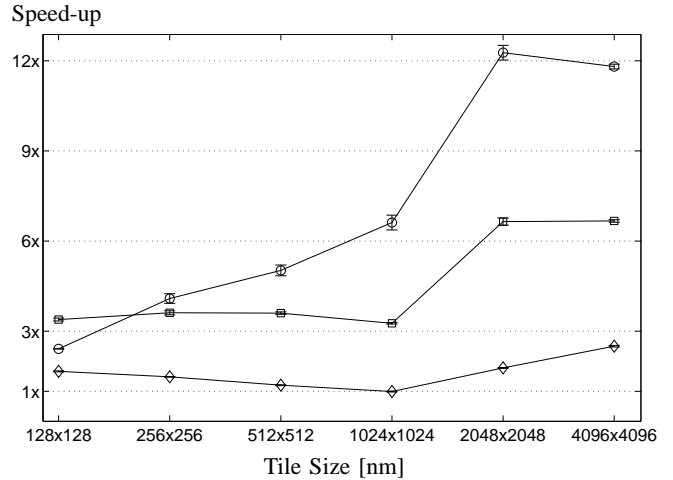}} \\
      \centerline{Tile Size [nm]} \\[-5pt]
    \end{minipage}
  }
  \caption{Average speed-up with 95\% confidence intervals. The three curves
           correspond respectively to the layers M1 ($\diamond$), M2 ($\scriptscriptstyle \square$) and CA ($\circ$).
           The speed-up is the ratio of the runtimes of the DHT and the PCHT.}
  \flabel{N_vs_SU}
\end{figure}





\bibliographystyle{IEEEbib}
\bibliography{IEEEabrv,bibliography}

\end{document}